\documentstyle[aps]{revtex}

\begin{document}


\title{Alpha Chain Structures of $^{12}$C}

\author{S. H. Hong\footnote{Electronic address: reds@physics.kyunghee.ac.kr}
and S. J. Lee\footnote{Electronic address: ssjlee@nms.kyunghee.ac.kr}}
\address{
Institute of Natural Sciences,
Kyung Hee University, Suwon, 449-701 }

\maketitle

\begin{abstract}
$N$-$\alpha$ structures of light nuclei with axial symmetry are studied
using relativistic Hartree approximation. Metastable excited states are
searched in a configuration space which allows linear alpha chain structures.
As a result, it is shown that $^{12}$C has $^8$Be$ + \alpha$ resonance
state at about 1 MeV above $^8$Be--$\alpha$ threshold as an asymmetric
3-$\alpha$ linear-chain structure, which plays an important role in stellar
nucleosynthesis.
\end{abstract}




Hydrogen burning and helium burning are very important processes
in the evolution of young stars, playing roles as the nuclear synthesizer
for light nuclei and as the source of energy production.
An interesting feature appearing in the helium burning process is
the possible $N$-$\alpha$ structures of light nuclei.
Since the even-even light nuclei, such as $^4$He, $^8$Be, $^{12}$C,
$^{16}$O, and $^{20}$Ne, have similar binding energies per nucleon but
larger than that of an alpha particle, the $^4$He nucleus in its ground
state, we may assume that these nuclei are composed of alpha particles
at least for some features of these nuclei.
\"Opik \cite{opik} and Salpeter \cite{salpt} pointed out that
unstable $^8$Be can capture additional $\alpha$ particle to
form $^{12}$C within their life time in a star, which is composed
of $^4$He as the ash of its hydrogen burn.
Most of $^{12}$C in our universe may be produced by
$^8$Be$ + \alpha \leftrightarrow ^{12}$C reaction.
Hoyle suggested \cite{hoyle} that this reaction must proceed through
a hypothetical $0^+$ resonance state of $^{12}$C at energy of 0.4 MeV
above 3-$\alpha$ threshold.
The second $0^+$ excited state of $^{12}$C was found with the resonance
energy of $E_r = 0.3796$ MeV above the threshold
(0.278 MeV above $^8$Be$+ \alpha$ threshold)
and the full width of $\Gamma=8.5 \times 10^{-6}$ MeV
\cite{ajzenberg,ajzenberg2}.

In the $N$-$\alpha$ structure model, $^{12}$C is considered to be made
up of three alpha particles bound to each other.
Three alpha particles form an equilateral triangle in its ground state
\cite{fedorov} and form a linear chain for the $0^+$ excited state.
Alpha structures of the low-lying states of $^{12}$C and $^{16}$O,
including their $0^+$ excited states, have been studied within a
framework of nonrelativistic macroscopic and microscopic methods
\cite{fedorov,visschers,csoto,hayes,itagaki,pichler,descou,dunbar,pauli}.
These studies treat alpha particles as inert clusters,
and thus miss any detailed structure of nuclei and any change of the
internal structure of alpha particles in a nucleus.
For the study of detailed nuclear structure in these
highly excited super-deformed nuclei,
we need to use a mean field approach.

In nuclear mean field theory, a nuclear system is composed of
nucleons interacting strongly through a mean field potential.
While the nonrelativistic mean field approach uses a phenomenological
mean field potential as a function of nucleon density,
the mean field potential in relativistic mean field theory
is determined as a resultant of meson exchange.
After Walecka proposed a relativistic mean field (RMF) approach
for a nuclear system in which Dirac nucleons interact
by exchanging classical meson fields \cite{walecka},
it has been extended to include various quantum effects and
successfully applied in describing nuclear matter and finite nuclei
\cite{walecka,serot-walecka,sjlee,deform,deform2,sjleeth,%
reinhard,jkpsnm,jkpspa,jkps,rbhf,rhf,jkpsb}.
However the usage of the mean field approach is limited to
spherical or moderately deformed nuclei
due to the limitation of the size of the functional basis space.
To describe highly deformed alpha chain structure,
we construct, in this paper, a functional basis space for the axially
symmetric RMF calculation by appropriately locating single particle
levels of $^4$He.


In the relativistic mean field approach, nucleon-nucleon nonlocal
interactions are replaced by nucleon-meson local interactions
\cite{serot-walecka} of various meson exchange.
For the long range attractive and short range repulsive characteristics
of nucleon-nucleon interaction, we at least include in RMF
an isoscalar scalar meson field ($\sigma$ meson) and
a massive isoscalar vector meson field ($\omega$ meson).
An isovector vector meson field ($\rho$ meson) is included to handle
the charge exchange and the electromagnetic field for the electromagnetic
interaction among protons.
We also include the nonlinear self-interaction terms of the scalar
meson field to improve the compressibility of nuclear matter and
the deformation of finite nuclei.
By choosing carefully the coupling constants and the meson masses
as the parameters,
we can get quantitative agreement with experiments for spherical and
deformed nuclei \cite{serot-walecka,sjlee,deform,deform2,sjleeth,%
reinhard,jkpsnm}.
Lee {\it et al.} studied deformed nuclei within the RMF with various
parameter sets \cite{sjlee}. We have used the same numerical
methods and parameter sets for our calculation here.


For an $\alpha$ particle, linear parameter sets L1, L2, and L3 give too
little binding energy (15.4 MeV $\sim$ 3.6 MeV) while the nonlinear
parameter set NL1 gives about the right binding energy (31.9 MeV)
compared to the empirical value of 28.27 MeV \cite{ajzenberg}.
On the other hand, for the ground states of $^{12}$C and $^{16}$O,
NL1 gives much stronger binding (136.1 MeV and 206.5 MeV respectively)
while L1 gives 94.1 MeV and 128.9 MeV compared to the
empirical values of 92.2 MeV and 127.6 MeV \cite{matt}.
These results indicate that we need to search for a better parameter
set which is good for both $^4$He and $^{12}$C to study the alpha
structure of $^{12}$C.
However, for simplicity, we used parameter set L1 in this paper
since we are interested more in the existence of the alpha-chain
structure of $^{12}$C within a self-consistent mean field theory
rather than the detailed properties.

The numerical method is exactly the same as in Ref.\cite{sjlee}
except the methods of how to set the functional basis space and the
initial configuration of nuclei for the self-consistent calculation.
Initially, $^4$He nuclei in its ground state are linearly arranged
on the $z$-axis and each $^4$He nucleus is accompanied by several
excited single nucleon levels of $^4$He forming a large enough functional
basis space to handle highly deformed alpha-chain structure.

Starting with two $\alpha$ particles placed 4 fm to 13 fm apart,
nucleons in the 2-$\alpha$ system reassembled to form the ground
state of $^8$Be, with the binding energy of 43.1 MeV (empirical
binding energy is 56.5 MeV).
This state has a peanut shaped prolate deformation
with 2.8 fm separation between the supposed two alpha particles.
This state is about 13 MeV lower than the two-alpha threshold
in contrast to the empirical value which shows that the ground
state of $^8$Be has almost the same energy
as the two-alpha threshold, thus making $^8$Be unstable.
Remember here that the L1 parameter set produces too little binding for
small nuclei such as He and Be, the worse for the smaller nuclei.
When the initial separation of alpha particles was above 13 fm,
we got an excited oblate state of binding energy 34.5 MeV for the L1 set.
These converged solutions are just the same as the original method
of Ref.\cite{sjlee} calculated without any consideration of alpha structure.

Starting with three $\alpha$ particles linearly arranged with
separations of 4 fm to 10 fm, three $\alpha$ particles combined into
the ground state of $^{12}$C with binding energy of 94.0 MeV.
For calculations with initial separations above 10 fm,
we got continuum states in which
two $\alpha$ particles combined together to form the ground state of $^8$Be
and the remaining $\alpha$ particle was separated
away from the $^8$Be on its symmetry axis.
Both the $^8$Be piece and the $^4$He piece of this system have the same
binding energies, rms radii, and quadrupole moments as the ground state
of $^8$Be and the ground state of $^4$He respectively.
The continuum state solutions show that $^{12}$C has a fission barrier
of $\alpha$ emission at around 10 fm separation.
The energy of this $^8$Be--$^4$He system is highest when the separation
is about 10 fm and the energy is lowered toward
the $^8$Be$ + \alpha$ threshold as the separation becomes larger.
At around 10 fm separation the energy is $-57.2$ MeV which
is 1.3 MeV above the $^8$Be$ + \alpha$ threshold ($-58.5$ MeV for L1 parameter).
Although the energy is much higher than the empirical
value ($-84.54$ MeV) for the second $0^+$ state of $^{12}$C,
we may relate this continuum region to the $0_2^+$ excited state.
Experimentally, the $0_2^+$ excited state of $^{12}$C is 0.38 MeV above the
three-alpha threshold and 0.28 MeV above the $^8$Be$ + \alpha$ threshold
\cite{ajzenberg,ajzenberg2}.
Remembering that the L1 parameter set gives too small binding energy
for light nuclei and good fit for $^{12}$C and $^{16}$O,
this $^8$Be--$^4$He system might actually be the $^8$Be$ + \alpha$
resonance which is an asymmetric three-$\alpha$ linear chain state in
the alpha cluster model. This state is not a three-alpha resonance
state having equally spaced $\alpha$ particles.
In these calculations no three-alpha resonance state was found.

We have also tested this resonance state starting with initial states
built with a ground state $^8$Be and a ground state $^4$He placed on
the symmetry axis of $^8$Be.
When the initial separation is 4 fm to 10 fm apart, the system converged
to the ground state of $^{12}$C and the system becomes a $^8$Be$ + \alpha$
resonance state for the initial separation of 10 fm to 20 fm.
We, within our calculations, have found no resonance state of the excited
oblate $^8$Be and $\alpha$ which may correspond to an isosceles triangular
configuration in an alpha cluster model.
Even if we started with the initial configuration built with the excited
oblate $^8$Be and an alpha, the calculation converged to the resonance
state of the ground state $^8$Be and an alpha particle for any
initial separation.


In this paper, we have studied the alpha structure of light nuclei
in a relativistic mean field approach.
As a result, we have seen that the ground state of $^8$Be
may be considered as two alpha clusters separated 2.8 fm apart with
some overlap of surface region.
We have also seen that $^{12}$C has a continuum state region
which might be related to a $^8$Be$ + \alpha$ resonance
state as the $0^+$ state which is empirically at 0.38 MeV above the
three $\alpha$ threshold and that there is no 3-$\alpha$ resonance state.
These results confirm that $^{12}$C might be produced
through $^8$Be$ + \alpha$ resonance state in a star rather than
a direct combination of three alpha
particles through a 3-$\alpha$ resonance.
To check if the continuum state actually corresponds to a resonance state,
we need to find resonance width or level density by extending our method
or combining with other method.
We may apply this calculation in study of $\alpha$ structure
in $^{16}$O, $^{20}$Ne, and $^{24}$Mg.
For the study of more detailed structure, we should first
find a better parameter set which is good for the mass range
of 4 to 30.

This work was supported in part by Korea Research Foundation through
Basic Science Research Institute Grant No. 98-2422
and in part by Kyung Hee University through the Institute of Natural
Sciences under Grant No. 2u01-98-024.


\vspace{1cm}
\begin{thebibliography}{99}

\bibitem{opik} G. K. \"Opik, Proc. Roy. Irish Acad. {\bf A54}, 49 (1951).

\bibitem{salpt} E. E. Salpeter, Phys. Rev. {\bf 88}, 547 (1952);
Astrophys. J. {\bf 115}, 326 (1952); Ann. Rev. Nucl. Sci.{\bf 2}, 41 (1953);
Phys, Rev, {\bf 107}, 516 (1957).

\bibitem{hoyle} F. Hoyle, D. N. F. Dunbar, W. A. Wenzel and
W. Whaling, Phys. Rev. {\bf 92}, 1095 (1953);
F. Hoyle, Astrophys. J. Suppl., 121 (1954).

\bibitem{ajzenberg} F. Ajzenberg-Selove, Nucl. Phys. {\bf A490}, 1 (1988).

\bibitem{ajzenberg2} F. Ajzenberg-Selove, Nucl. Phys. {\bf A560}, 64 (1990).

\bibitem{fedorov} D. V. Fedorov and A. S. Jensen, Phys. Lett. {\bf B389},
 631 (1996).

\bibitem{visschers} J. L. Visschers and R. van Wageningen, Phys. Lett.
 {\bf B34}, 455 (1977); M. Vallier\'es, H. T. Coelho and T. K. Das,
 Nucl. Phys. {\bf A271}, 95 (1976).


\bibitem{csoto} A. Cs\'ot\'o, Phys. Rev. {\bf C52}, 2809 (1995).

\bibitem{hayes} A. C. Hayes and S. M. Sterbenz, Phys. Rev. {\bf C52},
 2807 (1995).

\bibitem{itagaki} N. Itagaki, A. Ohnishi, K. Kato, Nucl-th/9606056,
 LANL Preprint.

\bibitem{pichler} R. Pichler, H. Oberhummer, A. Cs\'ot\'o, S. A. Moszkowski,
 Nucl. Phys. {\bf A618}, 55 (1997).

\bibitem{descou} P. Descouvemont and D. Baye, Phys. Rev. {\bf C36},
 54 (1987).

\bibitem{dunbar} D. N. F. Dunbar, R. F. Pixley, W. A. Wenzel and
 W. Whaling, Phys. Rev. {\bf 92}, 649 (1953).

\bibitem{pauli} R. T. Pauli, Arkiv. Fysik {\bf 9}, 571 (1955);
 K. Ahnlund, Arkiv. Fysik {\bf 10}, 369 (1956).

\bibitem{walecka} J. D. Walecka, Ann. Phys. (N.Y.) {\bf 83}, 491 (1974).

\bibitem{serot-walecka} B. D. Serot, J. D. Walecka, Adv. in Nucl.
 Phys. {\bf 16}, 1 (1985).

\bibitem{sjlee} S. J. Lee, J. Fink, A. B. Balantekin, M. R. Strayer,
 and A. S. Umar, P. -G. Reinhard, J. A. Maruhn and W. Greiner,
 Phy. Rev. Lett. {\bf 57}, 2916 (1986); {\bf 59}, 1171 (1987).

\bibitem{deform}C.E. Price and G.E. Walker, Phys. Rev. {\bf C36}, 354 (1987).

\bibitem{deform2}W. Pannert, P. Ring and J. Boguta, Phys. Rev. Lett. {\bf 59},
 2420 (1987).

\bibitem{sjleeth}S.J. Lee, Ph.D. dissertation, Yale University, 1986.

\bibitem{reinhard} P. G-. Reinhard, M. Rufa, J. Maruhn, W. Greiner, and J.
  Friedrich, Z. Phys. {\bf A323}, 13 (1986).

\bibitem{jkpsnm}J.H. Lee, Y.J. Lee, and S.J. Lee,
 J. Korean Phys. Soc. {\bf 30}, 169 (1997).

\bibitem{jkpspa}S.J. Lee, J. Korean Phys. Soc. {\bf 32}, 612 (1998).

\bibitem{jkps}Doohwan Lee, Namyoung Lee, Dongwoo Cha, Hyoung Chan Bhang,
  Jumg-Hwan Jun and Hong Jung, J. Korean Phys. Soc. {\bf 30}, 643 (1997).

\bibitem{rbhf}R. Machleidt, Adv. Nucl. Phys. {\bf 19}, 189 (1989).

\bibitem{rhf}H. F. Boersma and R. Malfliet, Phys. Rev. {\bf C49},
 233 (1994); {\bf C49}, 1495 (1994).

\bibitem{jkpsb}Ghi R. Shin, J. Korean Phys. Soc. {\bf 29}, 571 (1996).

\bibitem{matt} J. H. E. Mattauch, W. Thiele, and A. H. Wapstra, Nucl. Phys.
 {\bf 67}, 1 (1965).

\end{thebibliography}
\end{document}